\documentclass[balancelastpage,twocolumn,pre,amsmath,amssymb,showpacs]{revtex4}

\usepackage{graphicx}% Include figure files
\usepackage{dcolumn}% Align table columns on decimal point
\usepackage{bm}% bold math
\usepackage{epsf}
 
\begin{document}

\title{Oscillatory mean-field dynamos with spherically symmetric, 
isotropic $\alpha$}

\author{Frank Stefani}
\email{F.Stefani@fz-rossendorf.de}
\author{Gunter Gerbeth}
\email{G.Gerbeth@fz-rossendorf.de}
\affiliation{Forschungszentrum Rossendorf, P.O. Box 510119,
D-01314 Dresden, Germany}%
 
\date{\today}

\begin{abstract}
Until recently, the existence of oscillatory mean-field dynamos of the
$\alpha^2$-type
with spherically symmetric and isotropic $\alpha$ was an open 
question. We find such dynamos by means of an 
evolutionary strategy, and we illustrate the spectral properties
of the corresponding dynamo operators. 
\end{abstract}
\pacs{47.65.+a, 52.65.Kj, 91.25.Cw}  

%\tableofcontents
\maketitle

For decades, homogeneous dynamos have been the subject of purely 
theoretical research that tried to explain magnetic field 
generation in
such inaccessible regions as the Earth's deep interior, the 
sun's 
convection zone, or the spiral arms of galaxies.
In 1999, this situation has changed with the first 
successful dynamo
experiments at the sodium facilities in Riga  \cite{PRL1,PRL2} 
and Karlsruhe \cite{MUST,STMU}. Now, 
dynamos 
run in laboratory, and their kinematic 
behavior can be predicted with an
error margin of a few percent \cite{RMP}. Dynamo theory is not 
unaffected by those
developments in dynamo ``engineering''. 
Whereas in the past the relevance of a dynamo model was 
ultimately
judged by its applicability to real cosmic bodies, it is 
now sensible
to "tailor" particular models which could show interesting 
features, e.g. field reversals or mode switching, in the
laboratory. 

These developments give also new impetus on spectral
analysis of dynamo operators. One of the simplest 
dynamo  models is a mean-field dynamo with a spherically 
symmetric and
isotropic helical turbulence parameter $\alpha$. In particular, the case
with constant $\alpha$ is one of the rare models in dynamo theory that
can be solved semi-analytically \cite{KRST,KRRA}. 
Though the construction of a spherically symmetric, isotropic
$\alpha^2$-dynamo seems away from realistic cosmic 
dynamos with their,
for instance, typical antisymmetry with respect to the equatorial
mid-plane, the basic analysis of this simplest case 
of $\alpha^2$-dynamos
is a necessary starting point for the investigation
of more realistic dynamos.
Despite 
its simplicity, the spherically
symmetric, isotropic $\alpha^2$-dynamo exhibits a rich spectral 
structure 
if  $\alpha$ is allowed to vary with the radius \cite{STGE1,STGE2}. 
In particular, complex eigenvalues have been found for a number of
models \cite{RAE86,RABR,SCZH,STGE2}. However, 
it always turned out that 
other modes 
with higher degrees $l$ of the 
spherical harmonics had larger growth rates than
the considered oscillatory mode (cf. \cite{RUEO}).
An explicit example of a spherically symmetric, 
isotropic $\alpha^2$-dynamo 
with the dominating mode being oscillatory 
does not exist up to present.

Thus motivated, our concrete goal in the present paper
is to solve the following inverse problem:
Find, for an $\alpha^2$-dynamo with spherically symmetric, 
isotropic $\alpha$,
a radial dependence $\alpha(r)$ so that the eigenmode for $l=1$ 
has zero 
growth rate and  non-zero frequency
{\it and} that all other modes have negative growth rates.
This task fits into a class of recently treated inverse problems 
concerning the determination of
$\alpha(r)$ from the demand that certain spectral properties 
(a few given eigenvalues, \cite{STGE1,STGE2})
or spatial properties of the magnetic field 
(``hidden dynamos'', \cite{TILG})
are fulfilled.

In the following, we will shortly describe the forward solver 
that is used for the determination of the eigenvalues, and the
evolutionary strategy (ES) which we employ to solve the 
inverse problem
formulated above.

We start with
the  induction equation 
for a  mean-field dynamo model with a
spherically symmetric
$\alpha$-coefficient (for the basics 
of mean-field dynamos see \cite{KRRA}).  
Inside a sphere of radius $R$ 
the electrical conductivity $\sigma$ is  
constant, whereas it is zero in the exterior. 
In the interior of the sphere the 
magnetic field ${\bm{B}}$ 
has to satisfy the induction equation 
\begin{eqnarray}
\frac{\partial {\bm{B}}}{\partial t} =\nabla 
\times (\alpha {\bm{B}}) +
\frac{1}{\mu_0 \sigma} \Delta {\bm{B}}
\end{eqnarray}
and the source-free condition
\begin{eqnarray}
\nabla \cdot {\bm{B}}=0 \; .
\end{eqnarray}

At the boundary $r=R$, the magnetic field has to match 
continuously to a potential field in the exterior. 

As usual \cite{KRRA}, we decompose ${\bm{B}}$ into a 
poloidal and a toroidal part,
\begin{eqnarray}
{\bm{B}}=-\nabla \times ({\bm{r}} \times 
\nabla S)-{\bm{r}} \times 
\nabla T \; ,
 \end{eqnarray}
with the defining scalars $S$ and $T$ expanded 
in spherical harmonics
according to 
\begin{eqnarray}
S(r,\theta,\phi)&=&\sum_{l=1}^{\infty} \sum_{m=-l}^{l}
R \, s_l^m(r) Y_l^m(\theta,\phi) \exp{(\lambda_{l} t)} \; ,\\
T(r,\theta,\phi)&=&\sum_{l=1}^{\infty} \sum_{m=-l}^{l}
t_l^m(r) Y_l^m(\theta,\phi) \exp{(\lambda_{l} t)} \; .
\end{eqnarray}

For the remaining of the paper, we will 
measure the length in units of $R$, 
the time in units of 
$\mu_0 \sigma R^2$, and
the parameter $\alpha$ in units of $(\mu_0 \sigma R)^{-1}$.

Using Eqs. (3-5), the induction equation  (1)
can be transformed into the eigenvalue equation system:
\begin{eqnarray}
\lambda_{l,n} s_l&=&
\frac{1}{r}\frac{d^2}{d r^2}(r s_l)-\frac{l(l+1)}{r^2} s_l
+\alpha(r) t_l \; ,\\
\lambda_{l,n} t_l&=&
\frac{1}{r}\frac{d}{dr}\left( \frac{d}{dr}(r t_l)-\alpha(r)
\frac{d}{dr}(r s_l) \right) \nonumber\\ 
 &&-\frac{l(l+1)}{r^2} 
(t_l-\alpha(r)
s_l) \; .
\end{eqnarray}

In our particular case of a spherically symmetric, isotropic 
$\alpha$ there is no coupling between field modes differing in 
the degree $l$ of the spherical harmonics.
The order $m$ of the coefficients 
of the defining scalars has  been skipped 
as it does not show up in the equations.
The radial wavenumber
is labeled by the index $n$.

The boundary conditions at $r=1$ are as follows:
\begin{eqnarray}
\frac{d s_l}{dr}|_{r=1}+{(l+1)} s_l(1)=t_l(1)=0 \; .
\end{eqnarray} 

A shooting technique and a fifth order Runge-Kutta
method was used to solve this system numerically, 
utilizing and adapting standard routines
from Numerical Recipes \cite{PTVF}. This code has been 
validated
by comparison with exact results (available only 
for $\alpha=constant$),
and with the results of other codes, including an integral 
equation solver \cite{XUSG}. 

For the description of the  inverse spectral solver we will be 
very short as it has been described in detail  
in \cite{STGE1,STGE2}. We use an evolutionary strategy (ES), with 
a ``population'' of ``individuals''
(parameter vectors that describe the function $\alpha(r)$) evolving
in a ``fitness'' landscape
according to principles taken from biology. Finally the  
evolution is stopped when the population has gathered
in the (hopefully) global maximum of the fitness landscape.
Note that in every step of the evolution the eigenvalue equation system
(6-8) is
solved correctly, without any compromise with other functionals to 
be minimized. 

This evolutionary strategy is very robust, and it can easily be
used for the solution of different inverse spectral problems
\cite{STGE1,STGE2}. For the solution
of our task we define the fitness $F[\alpha]$ of a function 
$\alpha(r)$ according to:
\begin{eqnarray}
F[\alpha] &=& -(\Re (\lambda_{1,1}[\alpha]))^2 \nonumber \\
&&-(1+\epsilon+\tanh(\Im (\lambda_{1,1}[\alpha]) - f))^{-1}\nonumber \\
&&-(1+\epsilon+\tanh(-\Re (\lambda_{2,1}[\alpha])))^{-1}\nonumber\\
&&-(1+\epsilon+\tanh(-\Re (\lambda_{3,1}[\alpha])))^{-1}\; .
\end{eqnarray}
What is the  motivation for this choice of the fitness? 
The first term, to begin with, tries to keep the  growth rate 
$\Re (\lambda_{1,1}[\alpha])$ of the eigenmode with $(l=1|n=1)$ as close 
as possible to zero. The second term
makes the frequency $\Im (\lambda_{1,1}[\alpha])$ of this mode
repel from the parameter $f$ which will be considered as variable.
The third and the fourth term are chosen so as to force the 
growth rates for $(l=2|n=1)$ and $(l=3|n=1)$  to be less than zero. 
In a strict 
sense one should also demand the growth rates for all the
higher values of $l$ and $n$ to be less than zero, but this can 
also be checked
a-posteriori.
The small value $\epsilon$, which is only used to avoid numerical
overflows, is chosen as 10$^{-6}$.
\begin{figure}
\epsfxsize=8.0cm\epsfbox{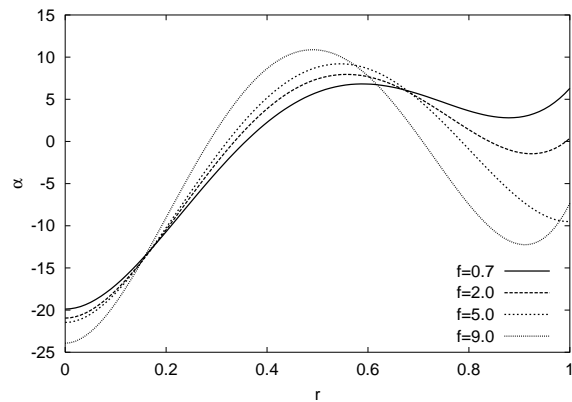}
\caption{The functions $\alpha(r)$ resulting from the evolutionary
strategy with fitness function $F[\alpha]$ according to 
Eq. (9) for four different values of $f$.}
\label{fig1}
\end{figure}

It should be mentioned that the ES can be combined with a certain
regularization of the functions $\alpha(r)$. By means of a
regularization 
parameter we can adjust the allowed mean quadratic curvature of $\alpha(r)$
\cite{STGE1}. It turns out that 
if we start with a large regularization parameter that
keeps the mean quadratic curvature of 
$\alpha(r)$ small we do not get any solution of our problem. Only
with a rather small regularization parameter that allows more curvature 
of $\alpha(r)$ the ES yields solutions.

In Fig. 1 we represent, for four different values of the parameter 
$f$, the 
functions $\alpha(r)$
resulting as  solutions of the ES. Note that the functions $\alpha(r)$ 
change 
their sign at least once. As we started with 
large regularization parameters without 
getting solutions, the curves in Fig. 1 can be considered
as the ``simplest'' ones (in the sense of minimal mean 
quadratic curvature) that 
fulfill our spectral demands. 
The analytical expressions for the
curves are given in the following table:\\

\begin{tabular}{cc}\hline\hline
$f$  & $\alpha(r)$ \\ \hline
0.7\;&-19.88+347.37\, $r^2$\; -656.71\, $r^3$+335.52\, $r^4$  \\
2.0\;&-20.95+399.40\, $r^2$\; -765.69\, $r^3$+387.61\, $r^4$  \\
5.0\;&-21.46+426.41\, $r^2$\; -806.73\,  $r^3$+392.28\,  $r^4$ \\
9.0\;&-23.93+594.35\, $r^2$-1243.02\,  $r^3$+665.27\,  
$r^4$ \\ \hline\hline
\end{tabular}\\\\

For these functions $\alpha(r)$, the
growth rates of the 
eigenmodes with $(l=1...3|n=1)$ 
and the frequency of the eigenmode with $(l=1|n=1)$  are as follows:\\

\begin{tabular}{ccccc}\hline\hline 
$f$&$\Re(\lambda_{1,1}[\alpha])$&$\Im(\lambda_{1,1}[\alpha])$&
  $\Re(\lambda_{2,1}[\alpha])$&$\Re(\lambda_{3,1}[\alpha])$ \\ \hline
0.7&0.00&4.98&-5.08&-12.02  \\
2.0&0.00&6.02&-5.35&-13.02 \\
5.0&0.01&7.24&-3.64&-10.82  \\
9.0&-0.11&9.30&-1.42&-9.32  \\ \hline\hline
\end{tabular}\\\\

These values have been
validated in an a-posteriori analysis of the spectrum for the 
functions $\alpha(r)$.

Evidently, the ES delivers functions $\alpha(r)$ that fulfill the
spectral demands formulated above. The $(l=1|n=1)$ eigenmode with zero 
growth rate is 
oscillatory, while the remaining eigenmodes  have growth 
rates less than 
zero. To be on the safe side, we have also 
computed the growth rates for
modes with $l>3$, which also turned out to be less than zero.
\begin{figure}
\epsfxsize=8.0cm\epsfbox{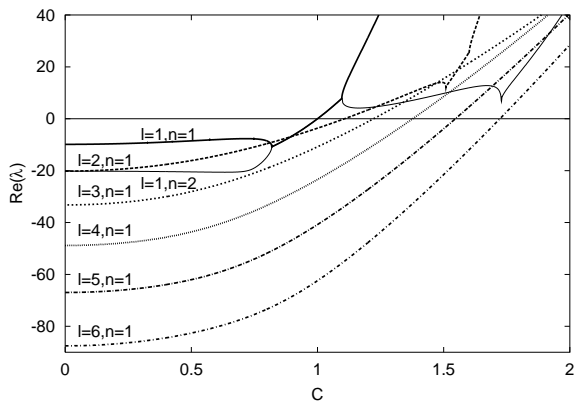}
\caption{Special case $f=5.0$. 
Growth rates for for the eigenfunctions with $(l=1|n=1,2)$ and
$(l=2...6|n=1)$.}
\label{fig2}
\end{figure}

\begin{figure}
\epsfxsize=8.0cm\epsfbox{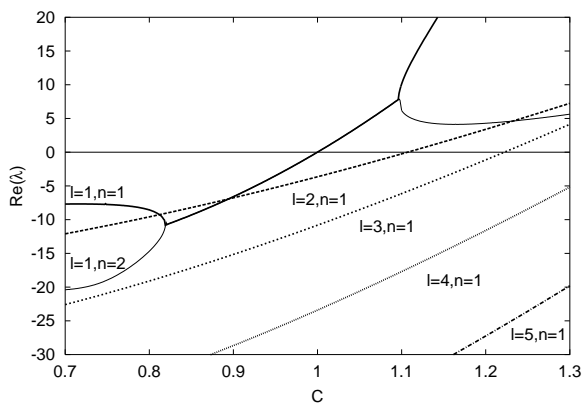}
\caption{Special case $f=5.0$. 
Growth rates for the eigenfunctions with $(l=1|n=1,2)$ and
$(l=2...5|n=1)$. Details close to the critical point $C=1$.}
\label{fig3}
\end{figure}

For the sake of illustration, let us concentrate in the 
following
on the spectral properties for the special case with $f=5$.
In Fig. \ref{fig2} we show the growth rates for the eigenmodes with
$(l=1|n=1,2)$ and $(l=2...6|n=1)$ as  functions of $C$,  
which is used to scale  the magnitude of 
$\alpha(r)$ (for $C=0$ we have the 
free decay case,
$C=1$ corresponds to the obtained functions $\alpha(r)$).
Details of this plot close to the critical point $C=1$ are
shown in Fig  \ref{fig3}.

At $C=0$, all modes start as non-oscillatory modes (with purely 
real eigenvalues).
At $C=0.818$, the two modes with $(l=1|n=1)$ and $(l=1|n=2)$ merge
and continue as a pair with complex conjugate eigenvalues 
(only a single line is shown). Interestingly enough, 
at $C=1.097$ this 
pair splits off again and the two modes with $(l=1|n=1)$ and $(l=1|n=2)$
continue separately, again with purely real eigenvalues. In between, at 
$C=1.00$ the pair crosses the abscissa, hence dynamo action occurs.
All modes with higher $l$ and $n$ have still negative growth rates 
at $C=1.00$.
\begin{figure}
\epsfxsize=8.0cm\epsfbox{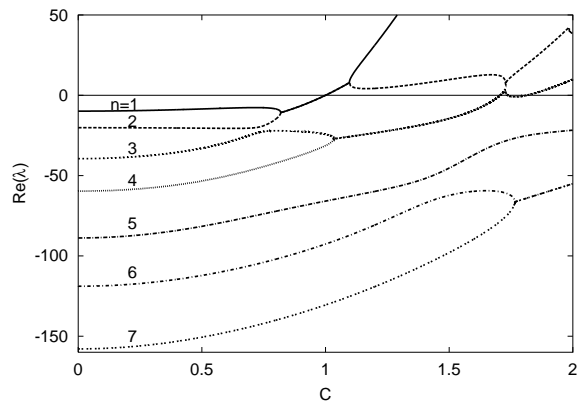}
\caption{Special case $f=5.0$. Growth rates for the 
eigenfunctions with $(l=1|n=1...7)$.}
\label{fig4}
\end{figure}
\begin{figure}
\epsfxsize=8.0cm\epsfbox{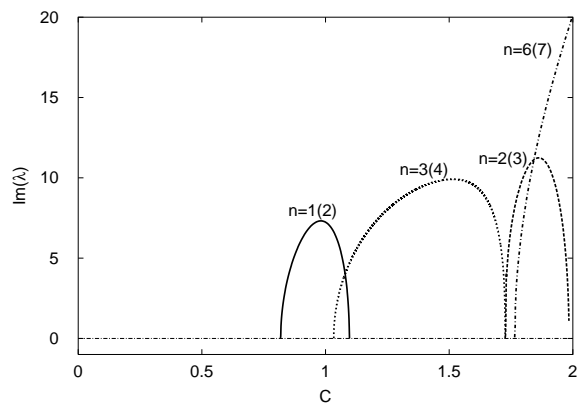}
\caption{Special case $f=5.0$. Frequencies 
for the eigenfunctions with $(l=1|n=1...7)$.}
\label{fig5}
\end{figure}

In Figs. \ref{fig4} and \ref{fig5} the growth 
rates and frequencies for the
modes with $(l=1|n=1...7)$ are  plotted. 
These figures give a flavor of the complexity of the spectrum, with
its merging and splitting points of modes with different $n$. 
\begin{figure}
\epsfxsize=8.0cm\epsfbox{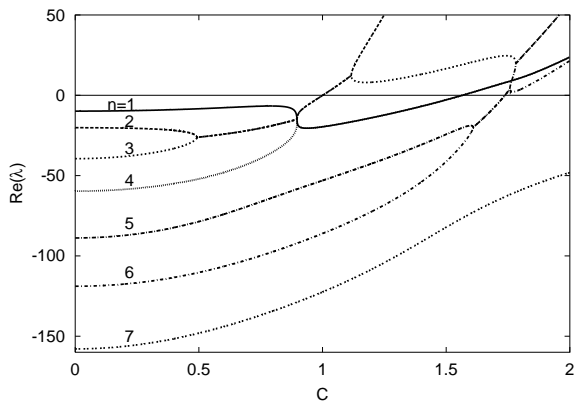}
\caption{Special case $f=9.0$. 
Growth rates for the eigenfunctions with $(l=1|n=1...7)$.}
\label{fig6}
\end{figure}

In this respect, another aspect should be addressed. Whereas the 
spectra for the functions $\alpha(r)$ for $f=0.7, 2.0, 5.0$ are 
rather similar, there is a transition if we go to $f=9.0$.
Figure \ref{fig6} shows the corresponding spectrum. 
Whereas in Fig. \ref{fig4}
we
observed a merging of the modes with $n=1$ and $n=2$, we get now 
a merging of the modes with $n=2$ and $n=3$. For larger $C$, the modes 
with $n=1$ and $n=4$ come together, meeting at the same point with 
the common line of the complex conjugated $n=2(3)$  modes.
\begin{figure}
\epsfxsize=8.0cm\epsfbox{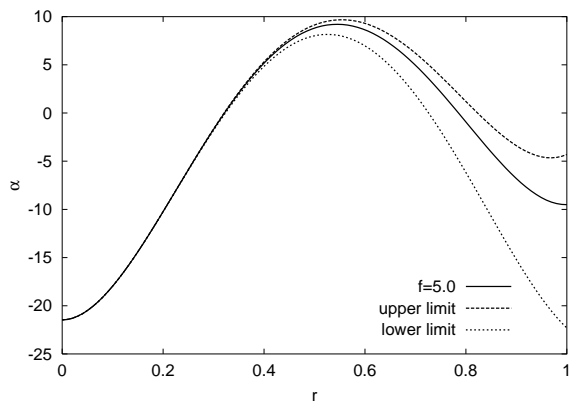}
\caption{Modified functions $\alpha(r)$ of the type obtained
for $f=5$, but 
with varying coefficient of the $r^4$ term, that are 
oscillatory dynamos.}
\label{fig7}
\end{figure}

Evidently, the dynamo operators connected with the obtained
functions $\alpha(r)$ show 
very interesting spectral features which warrants further analysis.

A complete characterization of 
the $\alpha(r)$-profiles that fulfill our spectral demands would require 
an extensive parameter study. At least we can get a certain feeling on the
broadness of  the "corridor" for a limited type of deformations. 
For this purpose, we start with the 
function $\alpha(r)=-21.46+426.41r^2\; -806.73 r^3+392.28 r^4$ which
was the solution for $f=5$. Now let us change the coefficient of 
the $r^4$ term,
leaving all other coefficients unchanged. Figure \ref{fig7} 
shows the upper and the 
lower limits of the corresponding deformations that still belong to our
oscillatory class of dynamos. Above the
upper limit, the $(l=2|n=1)$ mode starts to dominates. Below the lower 
limit, the $(l=4|n=1)$ mode 
dominates.

In summary, we have found a class of oscillatory mean-field dynamos 
with spherically symmetric, 
isotropic $\alpha$ working in a finite volume with 
homogeneous conductivity. The obtained functions $\alpha(r)$ are 
smooth and by no means exotic or of a pronounced layer type
\cite{RAE86,RABR}, but they are
characterized by at least 
one change of sign along the radius. 
In the sense of minimal 
averaged quadratic curvature, our solutions seem to be
the simplest ones.

As for the back-reaction regime, an interesting behaviour of
the considered dynamo models is conceivable. Even a slight 
modification of the functions $\alpha(r)$ due to the action 
of the Lorentz forces could trigger a transition to a 
non-oscillatory mode (either with $l=1$ or with higher $l$).

We do not claim any particular astrophysical
relevance of our result. However, keeping in mind that the
Karls\-ruhe dynamo experiment is an $\alpha^2$-dynamo (although 
with an anisotropic
$\alpha$-tensor) one could imagine a generalization of 
our method
to similar laboratory dynamos.

%\section*{ACKNOWLEDGMENTS}

\end{document}